\documentclass[fleqn,twoside,11pt]{article}
\usepackage{amssymb}
\usepackage{graphicx}
\usepackage{amsmath}
\begin{document}

%\noindent
%{\sf University of Shizuoka}

%\hspace*{13cm} {\large US-06-05}

\vspace{3mm}

\begin{center}
%\title{

{\Large\bf  S$_3$ Symmetry and Neutrino Masses and Mixings}

\vspace{3mm}
{\bf Yoshio Koide}

{\it Department of Physics, University of Shizuoka, 
52-1 Yada, Shizuoka 422-8526, Japan\\
E-mail address: koide@u-shizuoka-ken.ac.jp}

%\vspace{2mm}
\date{\today}
\end{center}

%\vspace{3mm}
%\maketitle
\begin{abstract}
Based on a universal seesaw mass matrix model with three
scalars $\phi_i$, and by assuming an S$_3$ flavor symmetry
for the Yukawa interactions, the lepton masses and
mixings are investigated systematically.
In order to understand the observed neutrino mixing, 
the charged leptons $(e, \mu, \tau)$ 
are regarded as the 3 objects $(e_1, e_2, e_3)$ of S$_3$, 
while the neutrino mass-eigenstates are regarded as
the irreducible representation $(\nu_\eta, \nu_\sigma, \nu_\pi$)
of S$_3$, where $(\nu_\pi, \nu_\eta)$ and $\nu_\sigma$ are a doublet
and a singlet, respectively, which are composed of the 3 objects
$(\nu_1, \nu_2, \nu_3)$ of S$_3$.
\end{abstract}

\vspace{3mm}
%%%%%%%%%%%%%%%%%%%%%%%%%%%%%%%%%%%%%%%%%%%%%%%%%%%%%%%%%%%%%

%%%%%%%%%%%%%%%%%%%%%%%%%%%%%%%%%%%%%%%%%%%%%%%%%%%%%%%%%%%%%
%\begin{multicols}{2}

{\large\bf 1 \ Introduction}

It is generally considered that masses and mixings of 
the quarks and leptons will obey a simple law of nature, 
so that we expect that 
we will find a beautiful relation among those values.
However, even if there is such a simple relation in the quark sector,
it is hard to see such a relation in the quark sector, because  
the relation will be spoiled by the gluon cloud.
We may expect that such a beautiful relation will be found just in
the lepton sector.
Therefore, in the present paper, we will confine ourselves to the 
investigation of the lepton masses and mixings.
Here, we would like to emphasize that we should search a model which
gives a reasonable description of not only the masses, but also 
the mixings.
Especially, we should direct our attention to the mixing pattern 
rather than to the mass spectrum in the neutrino sector.

It is also considered that the mass matrices of the fundamental 
particles will be governed by a symmetry.
In the present paper, we take notice of a permutation symmetry 
S$_3$ \cite{S3}.
Let us begin with giving a short review how useful a description
based on the S$_3$ symmetry is in the lepton masses and mixings.

The observed neutrino data have strongly suggested that the neutrino 
mixing is approximately described by the so-called tribimaximal mixing 
\cite{tribi}
$$
U_{TB}=\left(\begin{array}{ccc}
-\frac{2}{\sqrt6} & \frac{1}{\sqrt3} & 0 \\
\frac{1}{\sqrt6} & \frac{1}{\sqrt3} & -\frac{1}{\sqrt2} \\
\frac{1}{\sqrt6} & \frac{1}{\sqrt3} & \frac{1}{\sqrt2} \\
\end{array} \right) .
\eqno(1.1)
$$
The tribimaximal mixing is interpreted in the framework of S$_3$:
We define the doublet 
($\psi_\pi$,$\psi_\eta$) and singlet $\psi_\sigma$ of the permutation 
symmetry $S_3$ as
$$
\left(\begin{array}{l}
\psi_\pi \\
\psi_\eta \\
\psi_\sigma \\
\end{array} \right)=
\left(\begin{array}{ccc}
0 & -\frac{1}{\sqrt2} & \frac{1}{\sqrt2} \\
-\frac{2}{\sqrt6} & \frac{1}{\sqrt6} & \frac{1}{\sqrt6} \\
\frac{1}{\sqrt3} & \frac{1}{\sqrt3} & \frac{1}{\sqrt3} \\
\end{array} \right)
\left(\begin{array}{l}
\psi_1 \\
\psi_2 \\
\psi_3 \\
\end{array} \right) ,
\eqno(1.2)
$$
where $(\psi_1, \psi_2, \psi_3)$ are three objects of S$_3$. 
When we assume that the mass-eigenstates in the charged lepton sector are 
$(e_1,e_2,e_3)=(\tau,\mu,e)$, while those in the neutrino sector are 
$(\nu_\pi,\nu_\eta,\nu_\sigma)$ with the mass hierarchy 
$$
{m_{\nu\eta}}^2<{m_{\nu\sigma}}^2<{m_{\nu\pi}}^2 ,
\eqno(1.3)
$$
the neutrino mixing matrix $U_\nu$ of the basis 
$(\nu_\eta,\nu_\sigma,\nu_\pi)$ to the basis
$(e_1,e_2,e_3)=(e,\mu,\tau)$ is given by the form (1.1),
because the basis $(\nu_\eta,\nu_\sigma,\nu_\pi)$ is given by
$$
\left(
\begin{array}{l}
\nu_e \\
\nu_\mu \\
\nu_\tau \\
\end{array} \right) \equiv 
\left(
\begin{array}{l}
\nu_1 \\
\nu_2 \\
\nu_3 \\
\end{array} \right)=
\left(\begin{array}{ccc}
\frac{2}{\sqrt6} & \frac{1}{\sqrt3} & 0 \\
-\frac{1}{\sqrt6} & \frac{1}{\sqrt3} & -\frac{1}{\sqrt2} \\
-\frac{1}{\sqrt6} & \frac{1}{\sqrt3} & \frac{1}{\sqrt2} \\
\end{array} \right) 
\left(
\begin{array}{l}
\nu_\eta \\
\nu_\sigma \\
\nu_\pi \\
\end{array} \right) .
\eqno(1.4)
$$
Here, the weak iso-doublets are given by $(\nu_i, e_i)_L$
($i=1,2,3$) [and also $(\nu_a, e_a)_L$ ($a=\pi, \eta, \sigma$)].
In other words, in order to obtain the tribimaximal mixing, we must
build a model where the mass-eigenstates are 
$(e_1,e_2,e_3)=(\tau,\mu,e)$ and 
$(\nu_\pi,\nu_\eta,\nu_\sigma)$ with the mass hierarchy (1.3).

On the other hand, it is well-known that the observed charged 
lepton mass spectrum \cite{PDG06} satisfies the relation 
\cite{Koidemass,Koide90} 
$$
m_e+m_{\mu}+m_{\tau}=
\frac{2}{3}\left( \sqrt{m_e}+\sqrt{m_\mu}+\sqrt{m_{\tau}} 
\right)^2 , 
\eqno(1.5)
$$
with remarkable precision. The mass formula (1.5) is invariant under 
any exchange $\sqrt{m_i}\leftrightarrow\sqrt{m_j}$  ($i,j=e,\mu,\tau$).
This, too, suggests that a description by S$_3$ may be useful for 
a mass matrix model.

As an explanation of the mass formula (1.5), the author has proposed a model 
\cite{Koide90,KF96,KT96}
with 3 flavor scalars $\phi_i$ in the framework of the universal seesaw model
\cite{UnivSeesaw}: 
A fermion mass matrix $M_f$ is given by 
$$
M_f=m_L^fM_F^{-1}m_R^f ,
\eqno(1.6)
$$
where $M_F$ is a mass matrix of hypothetical heavy fermions $F_i$ 
$(i=1,2,3)$.
For example, for the charged lepton sector, we assume
$$
m_L^e=\frac{1}{\kappa}m_R^e=y_e{\rm diag}(v_1,v_2,v_3) ,
\eqno(1.7)
$$
($\kappa$ is a constant with $\kappa \gg 1$)
and $M_E\propto {\bf 1}\equiv {\rm diag}(1, 1, 1)$, where 
$v_i\equiv \langle\phi_{Li}^0\rangle = \langle\phi_{Ri}^0\rangle/\kappa$, 
and $m_L^e$ (and also $m_R^e$) is defined by 
$\bar{e}_L m_L^2 \rangle  E_R$.

If we assume that the vacuum expectation values (VEV) 
$v_i$ satisfy the relation 
$$
v_1^2 + v_2^2 + v_3^2 = \frac{2}{3}
 \left( v_1 + v_2 + v_3 \right) ^2 ,
\eqno(1.8)
$$
we can obtain the relation (1.5). 
Of course, here, we have assumed that the 
Yukawa interaction in the charged lepton sector is given by an $S_3$ 
invariant form 
$$
H_e = y_e \left( \bar{\ell}_{L1} \phi_{L1} E_{R1} 
+\bar{\ell}_{L2} \phi_{L2} E_{R2}+
\bar{\ell}_{L3} \phi_{L3} E_{R3} \right),
\eqno(1.9)
$$ 
(and also a similar interaction for $\bar{\ell}_R\phi_R E_L$),
where $\ell_{L/R}=(\nu_{L/R},e_{L/R})$, and 
$\phi_{L/R}=(\phi^+_{L/R}, \phi^0_{L/R})$. 
The form (1.9) is not a general form under the S$_3$ symmetry.
We have assumed the universality of the coupling constants in
additon to the S$_3$ symmetry.

The relation among the VEVs $v_i$, (1.8), can read 
$$
v_\pi^2+v_\eta^2=v_\sigma^2 ,
\eqno(1.10)
$$
in terms of S$_3$, 
because
$$
v_1^2+v_2^2+v_3^2=v_\pi^2+v_\eta^2+v_\sigma^2=2 v_\sigma^2=2
\left( \frac{ v_1+v_2+v_3}{\sqrt{3}} \right)^2 ,
\eqno(1.11)
$$
where $v_a =\langle \phi_a^0 \rangle$  ($a=\pi, \eta, \sigma$), and
$(\phi_\pi,\phi_\eta,\phi_\sigma)$ have been defined by Eq.(1.4). 
For a Higgs potential model based on an S$_3$ symmetry  which 
leads to the relation (1.10), for example, 
see Ref. \cite{Koide06}.
The $S_3$ symmetry is again related to the lepton masses and mixings.

Thus, it is likely that the $S_3$ symmetry (or a higher symmetry which include 
$S_3$) plays an essential role on a unified description of the lepton mass 
matrices. 
In the present paper, we will assume that, in the universal seesaw model 
with three scalars $\phi_i$, the Yukawa interactions are exactly invariant 
under the $S_3$ symmetry, 
and the $S_3$ symmetry is broken only by the VEVs $v_i$ of the 
three scalars $\phi_i$.  
For the seesaw mass matrix model (1.6), 
by inheriting the formulation in charged lepton sector, 
we assume as follows:
(i) $M_F$ have a unit matrix structure, 
at least, for the charged lepton and neutrino sectors, i.e.
$$
M_E \propto {\bf 1},  \ \ \ M_N \propto {\bf 1}.
\eqno(1.12)
$$
(ii) $m_L^f$ (and $m_R^f$) have sector-dependent ($f$-dependent)
structures.  We still assume the diagonal form
$$
m_L^e = y_e {\rm diag}(v_1^d, v_2^d, v_3^d) ,
\eqno(1.13)
$$
in the charged lepton sector, but
we consider that $m_L^\nu$ in the neutrino sector is not diagonal.
Therefore, in the present model (1.6), the neutrino mixing is 
caused by the structure of $m_L^\nu$.
We also assume that the VEVs $v_i^u$ satisfy
the relation (1.8) as well as $v_i^d$ in the charged lepton
sector. 
However, note that in spite of the assumption (1.8) for $v_i^u$,
the eigenvalues of the matrix $m_L^\nu$, in general, do not
satisfy a relation similar to Eq.~(1.8). 
The purpose of the present paper is to investigate what 
structure of the Dirac mass matrix $m_L^\nu$ in the seesaw mass 
matrix model (1.6) is required in order to fit the model for 
the neutrino oscillation data.

By the way, the seesaw-type model (1.6) with 3 scalars $\phi_{Li}$
(and $\phi_{Ri}$) causes some trouble, for example, the flavor changing
neutral currents (FCNC) problem, the spoiling of the asymptotic 
freedom of the SU(3) color, and so on.
Therefore, instead of the Yukawa interaction (1.9), we may consider 
a Frogatt-Nielsen 
\cite{Frogatt} type model with five dimensional operators
$\overline{\ell}_{Li} H_L \phi_{i} E_{Ri}$, 
where $H_L$ is the conventional SU(2)$_L$-doublet Higgs scalar
$H_L=(H_L^+, H_L^0)$, 
and $\phi_{i}$ are 3-family SU(2)$_L$-singlet scalars: 
$$
H_{eff}= y_e \bar{\ell}_L H_L^d  \frac{\phi^d}{\Lambda_d}
 E_R 
+y_\nu  \bar{\ell}_L H_L^u \frac{\phi^u}{\Lambda_u}
N_R  ,
\eqno(1.14)
$$
where  $\Lambda_f$ are scales of the effective theory.
We consider $m_W^2 = \frac{1}{2} g_w^2 (\langle H_d^0 \rangle^2
+\langle H_u^0 \rangle^2)$ and $\langle \phi^f \rangle /\Lambda_f
\sim 1$ ($f=u,d$).
Since we interest only in the flavor structure, for convenience, 
hereafter, we will drop the Higgs scalars $H_L^f$ from Eq.~(1.14) 
and we will call $\bar{\ell}_L H_L^u \phi^u N_R$  the Yukawa 
interaction $\bar{\ell}_L \phi^u N_R$ simply.

%%%%%%%%%%%%%%%%%%%%%%%%%%%%%%%%%%%%%%%%%%%%%%%%%%%%%%%%%%%%%%%%%%%%%%%%%%%%%%
\vspace{3mm}

{\large\bf 2 \ Mass eigenvalues}

In general, an $S_3$ invariant Yukawa interaction with 3 scalars 
$\phi_a$ $(a=\pi,\eta,\sigma)$ is given by 
$$
H=\left(
y_0\frac{\bar{\psi}_\pi \psi_\pi+\bar{\psi}_\eta \psi_\eta+\bar{\psi}_\sigma 
\psi_\sigma}{\sqrt3}+y_1\frac{\bar{\psi}_\pi \psi_\pi+\bar{\psi}_\eta 
\psi_\eta-2\bar{\psi}_\sigma \psi_\sigma}{\sqrt6}
\right) \phi_\sigma 
$$
$$
+y_2 \left(
\frac{\bar{\psi}_\pi \psi_\eta+\bar{\psi}_\eta \psi_\pi}{\sqrt2}\phi_\pi
+\frac{\bar{\psi}_\pi \psi_\pi-\bar{\psi}_\eta \psi_\eta}{\sqrt2}\phi_\eta
\right)
$$
$$
+y_3\frac{\bar{\psi}_\pi \phi_\pi+\bar{\psi}_\eta \phi_\eta}{\sqrt2}\psi_\sigma
+y_4\bar{\psi}_\sigma \frac{\phi_\pi \psi_\pi+\phi_\eta \psi_\eta}{\sqrt2} ,
\eqno(2.1)
$$
where we read $\bar{\psi}=\bar{\ell}_L \equiv (\bar{\nu_L},  \bar{e}_L)$ , 
$\psi=E_R$  and $\phi_a=\phi_a^d$ 
for the charged lepton sector, $\bar{\psi}=\bar{\ell}_L$, 
$\psi=N_R$ (or $\nu_R)$ and $\phi_a=\phi_a^u$ 
for the neutrino sector, and we have dropped $H_L^f/\Lambda_f$ 
for convenience. 
For example, the interaction (1.9) 
in the charged lepton sector 
corresponds to the case 
$$
y_0=y_e, \ \ y_1=0, \ \ y_2=\frac{1}{\sqrt3}y_e, \ \ y_3=y_4=
\sqrt{\frac{2}{3}}y_e .
\eqno(2.2)
$$
The Yukawa interaction (2.1) gives the mass matrix $m^f_L$ for the basis 
$(\psi_\pi,\psi_\eta,\psi_\sigma)$,
$$
m^f_L=
\left(\begin{array}{ccc}
\left(
\frac{y_0}{\sqrt3}+\frac{y_1}{\sqrt6}
\right) v_\sigma+\frac{y_2}{\sqrt2}v_\eta & 
\frac{y_2}{\sqrt2}v_\pi & \frac{y_3}{\sqrt2}v_\pi \\
\frac{y_2}{\sqrt2}v_\pi & 
\left(
\frac{y_0}{\sqrt3}+\frac{y_2}{\sqrt6}
\right) v_\sigma-\frac{y_2}{\sqrt2}v_\eta & 
\frac{y_3}{\sqrt2}v_\eta \\
\frac{y_4}{\sqrt2}v_\pi & \frac{y_4}{\sqrt2}v_\eta & 
\left(
\frac{y_0}{\sqrt3}-2\frac{y_1}{\sqrt6}
\right) v_\sigma \\
\end{array} \right) .
\eqno(2.3)
$$

Hereafter, for simplicity, we confine ourselves to investigating a case with a 
symmetric mass matrix form $(m^f_L)^T=m_L^f$, i.e. with $y_3=y_4$. 
Then, we have still 
5 parameters, $y_0v_\sigma, y_1v_\sigma, y_2v_\pi, y_3v_\pi$ and 
$v_\pi/v_\eta$, in the model, so that the model has no predictability. 
In the 
present paper, we do not impose a further symmetry on the model. 
Alternatively, we will investigate what constraints on the mass matrix 
parameters (or specific relations among those) are required from the 
phenomenological studies.

Now let us return to the subject on the neutrino Dirac mass matrix $m_L^\nu$ 
which is, in general, given by the form (2.3) on the basis 
$(\nu_\pi, \nu_\eta, \nu_\sigma)$. 
(Hereafter, for convenience, we will denote $y_i^\nu$ as $y_i$ simply.)
As we discussed in the previous section, 
the present neutrino 
oscillation data favor to the tribimaximal mixing, so that the neutrino states 
are approximately in the mass eigenstates $(\nu_\eta, \nu_\sigma, \nu_\pi)$ 
with $m_\eta^2<m_\sigma^2<m_\pi^2$. 
Therefore, for convenience, we investigate a case in the limit of $y_3=0$. 
(Since the observed neutrino mixing is not the exact tribimaximal mixing,
the condition $y_3=0$ is only an approximate requirement for convenience.)
The mass matrix with $y_3=0$ is diagonalized by a rotation 
$$
R(\theta_{\pi \eta})=\left(\begin{array}{ccc}
c_{\pi \eta} & s_{\pi \eta} & 0 \\
-s_{\pi \eta} & c_{\pi \eta} & 0 \\
0 & 0 & 1 \\
\end{array} \right) ,
\eqno(2.4)
$$
where $c_{\pi \eta}=\cos\theta_{\pi \eta}$ and 
$s_{\pi \eta}=\sin\theta_{\pi \eta}$, and 
$$
\tan2\theta_{\pi \eta}=-\frac{v_\pi}{v_\eta} ,
\eqno(2.5)
$$
as
$$
R^T(\theta_{\pi \eta}) m_L^\nu R(\theta_{\pi \eta})
={\rm diag}(m_\pi, m_\eta, m_\sigma) .
\eqno(2.6)
$$
The mass eigenvalues $m_\pi$, $m_\eta$ and $m_\sigma$ are given by 
$$
\begin{array}{lll}
m_\pi=\left(
\frac{y_0}{\sqrt3}+\frac{y_1}{\sqrt6}
\right)v_\sigma \pm \frac{|y_2|}{\sqrt2}\sqrt{v_\pi^2+v_\eta^2} , \\
m_\eta=\left(
\frac{y_0}{\sqrt3}+\frac{y_1}{\sqrt6}
\right)v_\sigma \mp \frac{|y_2|}{\sqrt2}\sqrt{v_\pi^2+v_\eta^2} , \\
m_\sigma=\left(
\frac{y_0}{\sqrt3}-2\frac{y_0}{\sqrt6}
\right)v_\sigma ,
\end{array} 
\eqno(2.7)
$$
where we have defined 
$$
\sqrt2y_0+y_1>0 ,
\eqno(2.8)
$$
and the upper and lower signs in $\pm |y_2|$ (and also $\mp |y_2|$) 
correspond to the cases $y_2v_\eta>0$ and $y_2v_\eta<0$, respectively.

In the previous section, we have assumed that the VEVs $v_i^d$ of the scalars 
$\phi_i^d$, which couple to the charged leptons, satisfy the relation (1.10). 
Therefore, we also assume that the VEVs $v_i^u$ of the scalar $\phi_i^u$, 
which couple to the neutrino sector, satisfy the relation
$$
(v_\pi^u)^2+(v_\eta^u)^2=(v_\sigma^u)^2\equiv\frac{1}{2}v_u^2 ,
\eqno(2.9)
$$
where we do not always consider 
$\langle\phi_i^u\rangle=\langle\phi_i^d\rangle$. 
Then, the mass eigenvalues (2.7) lead to 
$$
\begin{array}{lll}
m_\pi=\left(
\frac{1}{\sqrt6}y_0+\frac{1}{2\sqrt3}y_1\pm\frac{1}{2}|y_2|
\right)v_u , \\
m_\eta=\left(
\frac{1}{\sqrt6}y_0+\frac{1}{2\sqrt3}y_1\mp\frac{1}{2}|y_2|
\right)v_u , \\
m_\sigma=\left(
\frac{1}{\sqrt6}y_0-\frac{1}{\sqrt3}y_1
\right)v_u .
\end{array} 
\eqno(2.10)
$$
Note that the mass spectrum is independent of the parameters 
$v_\pi^u/v_\sigma^u$ and $v_\eta^u/v_\sigma^u$, and only depends on the 
parameters $y_1/y_0$ and $|y_2|/y_0$. 
On the other hand, as seen in Eq.(2.5), 
the mixing angle $\theta_{\pi\eta}$ is independent of the parameters 
$y_i$ and only 
depends on the parameter $v_\pi^u/v_\eta^u$.

As we discussed in Sec.1, the observed tribimaximal mixing suggests that the 
neutrino mass eigenstates are $(\nu_\eta,\nu_\sigma,\nu_\pi)$. If the mass 
hierarchy is a normal type, it demands $m_\eta^2<m_\sigma^2 \ll m_\pi^2$, and 
if it is an inverse type, it demands $m_\pi^2 \ll m_\eta^2<m_\sigma^2$. 
The conditions for $m_\eta^2<m_\sigma^2<m_\pi^2$ and 
$m_\pi^2<m_\eta^2<m_\sigma^2$ are given in Appendix.

By the way, we have still two adjustable parameters $y_1/y_0$ and $y_2/y_0$ 
to predict the neutrino mass spectrum. 
In the following sections, we will 
investigate two typical cases by putting assumptions for the coupling 
constants $y_0$, $y_1$ and $y_2$.
Of course, the assumptions must also be applicable to the charged lepton
coupling constants (2.2).

%%%%%%%%%%%%%%%%%%%%%%%%%%%%%%%%%%%%%%%%%%%%%%%%%%%%%%%%%%%%%%%%%%%%%%%%%%
\vspace{3mm}

{\large\bf 3 \ Case with $y_0^2=y_1^2+y_2^2$}

In the mass matrix (2.3), the $y_1$- and $y_2$-terms are traceless, while the 
trace of the $y_0$-term is not zero. This suggests that the $y_0$-term may be 
distinguished from the other terms under a higher symmetry. 
Therefore, by way of 
trial, we put the following normalization condition for the coupling constants
$$
y_0^2=y_1^2+y_2^2 +y_3^2 ,
\eqno(3.1)
$$
which is satisfied by the coupling constants (2.2) in the charged lepton 
sector. Since we have assumed that $y_3=0$ in the neutrino sector, 
we can explicitly write the condition (3.1) as
$$
y_1=y_0\sin\alpha, \ \ y_2=y_0\cos\alpha .
\eqno(3.2)
$$
Then, we can rewrite Eqs.(2.10) as
$$
\begin{array}{lll}
m_\pi=\left[
\frac{1}{\sqrt6}-\frac{1}{\sqrt3}\sin\left(
\alpha \mp \frac{2}{3}\pi
\right) \right] y_0v_u , \\
m_\eta=\left[
\frac{1}{\sqrt6}-\frac{1}{\sqrt3}\sin\left(
\alpha \pm \frac{2}{3}\pi
\right) \right] y_0v_u , \\
m_\sigma=\left[
\frac{1}{\sqrt6}-\frac{1}{\sqrt3}\sin\alpha
\right] y_0v_u ,
\end{array}
\eqno(3.3)
$$
where $-\frac{\pi}{2}\leq \alpha \leq \frac{\pi}{2} \ \ (\cos\alpha>0)$, and, 
for $\frac{\pi}{2}\leq \alpha <\frac{3}{2}\pi$, we substitute $\pi-\alpha$ for 
$\alpha$ in Eq.(3.3).

Note that the case with the condition (3.1) which leads to Eq.(3.3) 
gives the relation
$$
m_\pi^2+m_\eta^2+m_\sigma^2=\frac{2}{3}(m_\pi+m_\eta+m_\sigma)^2 .
\eqno(3.4)
$$
Since these masses $(m_\pi,m_\eta,m_\sigma)$ are Dirac masses 
in the neutrino seesaw mass matrix $M_\nu=m_L^\nu M_N^{-1}(m_L^\nu)^T$, 
if we take the heavy 
Majorana mass matrix $M_N$ with the unit matrix form, we obtain the neutrino 
masses which are proportional to $m_\pi^2$, $m_\eta^2$ and $m_\sigma^2$, 
respectively. 
Therefore, the neutrino masses will satisfy a relation similar to 
the charged lepton mass relation (1.1) .

The differences among $m_\pi^2$, $m_\eta^2$ and $m_\sigma^2$ are given
as follows:
$$
m_\pi^2-m_\eta^2=\pm\frac{1}{\sqrt3}\cos\alpha(\sqrt2+\sin\alpha)y_0^2v_u^2 ,
\eqno(3.5)
$$
$$
m_\pi^2-m_\sigma^2=\pm\frac{1}{\sqrt3}\cos\left(\alpha\mp\frac{\pi}{3}\right)
\left[\sqrt2-\sin\left(\alpha\mp\frac{\pi}{\sqrt3}\right) \right]
 y_0^2v_u^2 ,
\eqno(3.6)
$$
$$
m_\eta^2-m_\sigma^2=\mp\frac{1}{\sqrt3}\cos\left(\alpha\pm\frac{\pi}{3}\right)
\left[\sqrt2-\sin\left(\alpha\pm\frac{\pi}{3}\right) \right]
 y_0^2v_u^2 ,
\eqno(3.7)
$$
where $|\alpha|<\pi/2$. 
For a case with a normal hierarchy, 
we should read the upper signs in Eqs.(3.5)-(3.7), 
so that we obtain
$$m_\eta^2<m_\sigma^2<m_\pi^2 \ \ {\rm for} \ \ 
-\frac{\pi}{6}<\alpha<\frac{\pi}{6}.
\eqno(3.8)
$$
For a case with an inverse hierarchy, since we should read the lower signs 
in Eqs.(3.5)-(3.7), we obtain
$$m_\pi^2<m_\eta^2<m_\sigma^2 \ \ {\rm for} \ \ 
-\frac{\pi}{2}<\alpha<-\frac{\pi}{6}.
\eqno(3.9)
$$

Next, let us seek for the numerical value of $\alpha$ which gives 
the ratio of the observed values 
$\Delta m^2_{solar}=(7.9_{-0.5}^{+0.6})\times10^{-5}$ eV$^2$ \cite{solar} to 
$\Delta m^2_{atm}=(2.74_{-0.26}^{+0.44})\times10^{-3}$ eV$^2$ \cite{atm},
$$
R_{obs} \equiv \frac{\Delta m^2_{solar}}{\Delta m^2_{atm}}
=(2.9\pm0.5)\times10^{-2}.
\eqno(3.10)
$$
The predicted ratio $R(\alpha)$ is given by
$$
R(\alpha)\equiv \frac{m_\sigma^4(\alpha)-m_\eta^4(\alpha)}
{m_\pi^4(\alpha)-m_\sigma^4(\alpha)} ,
\eqno(3.11)
$$
for a normal hierarchy $m_\eta^2 <m_\sigma^2 \ll m_\pi^2$.
From $R(\alpha)=R_{obs}$, 
we find
$$
\alpha=\left(3.0_{+1.4}^{-1.2}\right)^\circ ,
\eqno(3.12)
$$
where the sign $\mp$ corresponds to the sign $\pm$ of the
experimental error in Eq.(3.10). 
Similarly, we seek for the case with an inverse hierarchy,
but, we find that there is no solution with 
an inverse hierarchy.

The solution $\alpha=({3.0}_{+1.4}^{-1.2})^\circ$ gives 
$$
\begin{array}{lll}
m_\eta=-(0.076_{-0.008}^{+0.006})y_0v_u , \\
m_\sigma=(0.38 \pm 0.01)y_0v_u , \\
m_\pi=(0.923 \mp 0.006)y_0v_u . 
\end{array}
\eqno(3.13)
$$
The result $m_\eta<0$ leads to the change of the sign 
$\sqrt{m_{\nu 1}} \rightarrow -\sqrt{m_{\nu 1}}$ in 
a relation similar to Eq.(1.5):
$$
m_{\nu 1}+m_{\nu 2}+m_{\nu 3}=
\frac{2}{3}\left( -\sqrt{m_{\nu 1}}+\sqrt{m_{\nu 2}}+\sqrt{m_{\nu 3}} 
\right)^2 . 
\eqno(3.14)
$$
The relation (3.14) for the neutrino masses has recently speculated 
by Brannen \cite{Brannen} based on an algebraic method (however,
the algebraic method is highly mathematical, and 
the physical meaning of the method is somewhat not clear in the ``masses
and mixings").

The values (3.13) predicts the following neutrino masses 
$$
\begin{array}{lll}
m_{\nu 1}=(3.5 \pm 0.5)\times10^{-4}\ {\rm eV} , \\
m_{\nu 2}=(8.7 \pm 0.2)\times10^{-3}\ {\rm eV} , \\
m_{\nu 3}=(5.23_{+0.40}^{-0.25})\times10^{-2}\ {\rm eV} ,
\end{array}
\eqno(3.15)
$$
from the input value $m_{\nu 3} =\sqrt{\Delta m^2_{atm}}$.

Generally, the masses $m_{fi}$ which satisfy the relation (1.5) 
[or (3.14)] are expressed by a bilinear form
$$
m_{fi} = (z_{fi})^2 m_{f0} ,
\eqno(3.16)
$$
where the sector-dependent parameters $z_{fi}$ are normalized as 
$(z_{f1})^2+(z_{f2})^2+(z_{f3})^2=1$.
Then, the parameters $z_{fi}$ can always be expressed by the form
$$
\begin{array}{lll}
z_{f1}=\frac{1}{\sqrt6}-\frac{1}{\sqrt3}\sin\xi_f , \\
z_{f2}=\frac{1}{\sqrt6}-\frac{1}{\sqrt3}\sin(\xi_f+\frac{2}{3}\pi) , \\
z_{f3}=\frac{1}{\sqrt6}-\frac{1}{\sqrt3}\sin(\xi_f+\frac{4}{3}\pi) ,
\end{array}
\eqno(3.17)
$$
where we have taken $z_{f1}^2<z_{f2}^2<z_{f3}^2$.
From the observed charged lepton mass values \cite{PDG06}, 
we obtain the numerical value of $\xi_e$
$$
\xi_e=\frac{\pi}{4}-\varepsilon=42.7324^\circ \ \ 
(\varepsilon=2.2676^\circ) .
\eqno(3.18)
$$
Note that, in the limit of $\varepsilon \rightarrow 0$, the
electron mass becomes zero.
We consider that the parameter $\varepsilon$ is a fundamental
parameter which governs the charged lepton mass spectrum.

Comparing the expression (3.3) (with the upper signs) with the expression 
(3.17), we find that the parameter $\alpha$ is connected to $\xi_\nu$ by 
the relation
$$
\alpha=\frac{\pi}{3}-\xi_\nu .
\eqno(3.19)
$$
Therefore, we obtain 
$$
\xi_\nu-\xi_e=\left(\frac{\pi}{3}-\alpha \right) -\left(\frac{\pi}{4}
-\varepsilon \right) 
=\frac{\pi}{12}+\varepsilon-\alpha .
\eqno(3.20)
$$
Since the value of $\alpha$, (3.12), which is a solution of 
$R(\alpha)=R_{obs}$, 
is very close to the value $\varepsilon=2.27^\circ$, (3.18), 
from the observed charged 
lepton masses, we can regard $\alpha$ as $\alpha=\varepsilon$. 
Then, we obtain a phenomenological relation
$$
\xi_\nu=\xi_e+\frac{\pi}{12} .
\eqno(3.21)
$$
The relation (3.21) has also been speculated by Brannen \cite{Brannen},
but the reason is still controversial.
(Of course, in the present model, there is no theoretical reason 
for $\alpha=\varepsilon$.)

%Note that in the present model with $y_3=y_4$, as discussed in 
%Sec.2, the mass spectrum of the neutrinos is independent of the VEVs $v_i^u$, 
%and it depends only on the values $y_0$, $y_1$ and $y_2$. On the other hand, 
%the charged lepton mass spectrum depends only on the VEVs $v_i^d$, because we 
%have assumed the universality of the Yukawa coupling constants. The parameter 
%$\xi_e$ (therefore, $\varepsilon$) is one which characterizes the VEV 
%spectrum $(v_1^d, v_2^d, v_3^d)$, while the parameters $\xi_\nu$ 
%(therefore, $\alpha$) in the present model is one which characterizes the 
%structure of the neutrino Yukawa coupling constants. Therefore, the parameter 
%$\xi_\nu$ is different in kind from the parameter $\xi_e$. At present, 
%it is an open question whether the coincidence $\alpha \simeq \varepsilon$ is 
%accidental or not.

%%%%%%%%%%%%%%%%%%%%%%%%%%%%%%%%%%%%%%%%%%%%%%%%%%%%%%%%%%%%%%%%%%%%%%%%%%%%%%
\vspace{3mm}

{\large\bf 4 \ Case with $y_0^2+y_1^2=y_2^2$}

In the previous section, we have assumed a constraint (3.1) on the Yukawa 
coupling constants $y_0$, $y_1$ and $y_2$. However, the theoretical basis of 
the constraint is not clear. In the present section, instead of the constraint 
(3.1), we assume another constraint
$$
y_0^2+y_1^2=y_2^2 +y_3^2 ,
\eqno(4.1)
$$
which is again satisfied by the Yukawa coupling constants (2.2) in the charged 
lepton sector. The condition (4.1) means a requirement of the universality of 
the coupling constants in an extended meaning: the coupling constants of 
$\bar{\psi}_a \psi_a$ ($a=\pi, \eta, \sigma$) to the scalars 
are normalized with the equal weights for the scalars 
$\phi_\sigma$ and $\phi_\pi$ ($\phi_\eta$).

In the neutrino sector, since we have assumed $y_3=0$, we can denote the 
condition (4.1) as
$$
y_0=y_2\cos\beta , \ \ y_1=y_2\sin\beta .
\eqno(4.2)
$$
Then, the mass eigenvalues (2.10) are expressed as follows :
$$
\begin{array}{lll}
m_\pi=\frac{1}{2}
\left[\sin(\beta+\phi_0)\pm1
\right]|y_2|v_u , \\
m_\eta=\frac{1}{2}
\left[\sin(\beta+\phi_0)\mp1
\right]|y_2|v_u , \\
m_\sigma=\frac{1}{\sqrt2}\cos(\beta+\phi_0)y_2v_u ,
\end{array}
\eqno(4.3)
$$
where
$$
\sin\phi_0=\sqrt{\frac{2}{3}} , \ \ \cos\phi_0=\frac{1}{\sqrt3} , 
\ \ (\phi_0=54.74^\circ) ,
\eqno(4.4)
$$
we have again taken the condition (2.8), i.e.
$$
\sin(\beta+\phi_0)>0 \ \ (-\phi_0<\beta<\pi-\phi_0) , 
\eqno(4.5)
$$
and the upper and lower signs in Eq.(4.3) correspond to the cases 
$y_2v_\eta>0$ (a normal hierarchy case) and $y_2v_\eta<0$ 
(an inverse hierarchy case), respectively.

From the expression (4.3), we find
$$
m_\pi^2+m_\eta^2+m_\sigma^2=y_2^2v_u^2 , 
\eqno(4.6)
$$
$$
m_\eta+m_\sigma+m_\pi=\sqrt{\frac{3}{2}}y_2v_u\cos\beta .
\eqno(4.7)
$$
Therefore, we obtain 
$$
\frac{\frac{2}{3}(m_\pi+m_\eta+m_\sigma)^2}{m_\pi^2+m_\eta^2+m_\sigma^2}
=\cos^2\beta=1-\sin^2\beta .
\eqno(4.8)
$$
Thus, the parameter $\beta$ in the present model denotes a deviation from 
the mass formula (3.14) [(3.4)].

Note that if we find a solution $\beta=\beta_1$ which gives 
$R(\beta)=R_{obs}$ 
[$R(\beta)$ is given by Eq.(3.11) with $\alpha \rightarrow \beta$, 
and $R_{obs}$ 
is given by Eq.(3.10)], the value $\beta_2= 2\phi_0-\beta_1$ 
[$\phi_0$ is defined by Eq.(4.4)] is also a solution of $R(\beta)=R_{obs}$. 
From the expression (4.3), it is obvious that the solutions 
$\beta_1$ and $\beta_2$ 
give the same values for $m_\pi$ and $m_\eta$, but they give the values with 
the opposite signs to each other for $m_\sigma$. We list those 
solutions of $R(\beta)=R_{obs}$ in Table 1, together with the values 
of $m_\eta$, $m_\sigma$ and $m_\pi$. 

In Table 1, we also list the predicted values of the neutrino masses 
$m_{\nu 1}=m_\eta^2/M_N$, $m_{\nu 2}=m_\sigma^2/M_N$ and 
$m_{\nu 3}=m_\pi^2/M_N$ ($M_N$ is a Majorana mass 
$M_N \equiv M_{N1}=M_{N2}=M_{N3}$ of the heavy 
neutrinos $N_i$). Here, as the input value, we have used 
$m_{\nu 3}=\sqrt{\Delta m_{atm}^2}=0.0523$ eV for the normal hierarchy case, 
and $m_{\nu 2}=\sqrt{\Delta m_{atm}^2}=0.0523$ eV for the inverse hierarchy 
case. 
At present, the 
numerical values of $m_{\nu i}$ should not be taken rigidly. 
Therefore, we 
have omitted the error values from Table 1.

\vspace{3mm}

{\large\bf 5 \ Neutrino mixing matrix}

As we discussed in Sec.2, the additional rotation 
$R(\theta_{\pi\eta})$ from the 
tribimaximal mixing, (2.4), depends only on the value $v_\pi^u/v_\eta^u$, 
and it is independent of the values of $y_0$, $y_1$ and $y_2$. 
In order to see the effects of the additional rotation $R(\theta_{\pi \eta})$, 
we change 
from the basis $(\nu_\pi, \nu_\eta, \nu_\sigma)$ defined by Eq.(1.4) into the 
basis $(\nu_\eta, \nu_\sigma, \nu_\pi)$ given by 
$$
\left(\begin{array}{l}
\nu_\eta \\
\nu_\sigma \\
\nu_\pi \\
\end{array} \right)=
U_{TB}^T\left(\begin{array}{l}
\nu_e \\
\nu_\mu \\
\nu_\tau \\
\end{array} \right) ,
\eqno(5.1)
$$
where $U_{TB}$ is the tribimaximal mixing matrix defined by Eq.(1.1). 
If $v_\pi/\nu_\eta \neq 0$, i.e. $R(\theta_{\pi \eta}) \neq {\bf 1}$, 
the neutrino mixing matrix $U_\nu$ is given by
$$
U_\nu=U_{TB}
\left(\begin{array}{ccc}
c_{\pi \eta} & 0 & s_{\pi \eta} \\
0 & 1 & 0 \\
-s_{\pi \eta} & 0 & c_{\pi \eta} \\
\end{array} \right)=
\left(\begin{array}{ccc}
-\frac{2}{\sqrt6}c_{\pi \eta} & \frac{1}{\sqrt3} & 
-\frac{2}{\sqrt6}s_{\pi \eta} \\
\frac{1}{\sqrt6}c_{\pi \eta}+\frac{1}{\sqrt2}s_{\pi \eta} & 
\frac{1}{\sqrt3} & \frac{1}{\sqrt6}s_{\pi \eta}
-\frac{1}{\sqrt2}c_{\pi \eta} \\
\frac{1}{\sqrt6}c_{\pi \eta}-\frac{1}{\sqrt2}s_{\pi \eta} & 
\frac{1}{\sqrt3} & \frac{1}{\sqrt6}s_{\pi \eta}
+\frac{1}{\sqrt2}c_{\pi \eta} \\
\end{array} \right) ,
\eqno(5.2)
$$
where $s_{\pi\eta}=\sin\theta_{\pi\eta}$ and 
$c_{\pi\eta}=\cos\theta_{\pi\eta}$,
i.e.
$$
\tan^2\theta_{solar}=\frac{1}{2c_{\pi \eta}^2} ,
\eqno(5.3)
$$
$$
\sin^22\theta_{atm}=\left(1-\frac{4}{3}s_{\pi \eta}^2\right)^2 ,
\eqno(5.4)
$$
$$
(U_\nu)_{13}^2=\frac{2}{3}s_{\pi \eta}^2 .
\eqno(5.5)
$$

For convenience, we define the following $z_i$-parameters
$$
\langle \phi_i^u \rangle=z_i^u v_u , \ \ 
\langle \phi_i^d \rangle=z_i^d v_d ,
\eqno(5.6)
$$
with the normalizations $\sum_i(z_i^u)^2=\sum_i(z_i^d)^2=1$.
Here, note that in Eq.~(3.10), we have already define the 
$z_{fi}$-parameters similar to the present $z_i^u$- and 
$z_i^d$-parameters. 
In the charged lepton sector, since $\sqrt{m_{ei}}\propto v_i^d$,
the relation $z_{ei}=z_i^d$ holds.
However, in the neutrino sector, since $m_L^\nu$ is not diagonal,
the values $z_{\nu i}$ are not identical with $z_i^u$.

For the $z_i^d$-parameters, from the relation (1.5), we obtain 
$$
\frac{z_1^d}{\sqrt{m_e}}=\frac{z_2^d}{\sqrt{m_\mu}}=\frac{z_3^d}{\sqrt{m_\tau}}
=\frac{1}{\sqrt{m_e+m_\mu+m_\tau}} ,
\eqno(5.7)
$$
i.e. 
$$
z_1^d=0.016473, \ \ z_2^d=0.23678, \ \ z_3^d=0.97140 .
\eqno(5.8)
$$

If we assume $z_i^u=z_i^d$, 
we obtain $z_\pi^u=0.51939$, 
$z_\eta^u=0.47982$ and $z_\sigma^u=1/\sqrt2$ from the definition (1.4). 
Then, the rotation angle 
$\theta_{\pi\eta}=-(1/2)\tan^{-1}(v_\pi^u/v_\eta^u)=-23.63^\circ$ is 
too large to explain the observed neutrino mixings [see Eqs.(5.3)-(5.5)],
so that the case $z_i^u=z_i^d$ is ruled out. 

By the way, it is well known that the so-called
$2\leftrightarrow 3$ symmetry \cite{23sym} is  promising 
for neutrino mass matrix description.
Therefore, the simplest assumption is to require the 
$2\leftrightarrow 3$ symmetry for the VEV values
$\langle \phi_i^u \rangle$, i.e. $v_2^u=v_3^u$, which leads to
$$
v_\pi^u = 0 .
\eqno(5.9)
$$
The case gives $\theta_{\pi \eta} =0$ from Eq.~(2.5), so that
the neutrino mixing is exactly given by the tribimaximal
mixing (1.1).
Note that if we required the $2\leftrightarrow 3$ symmetry for
the fields $\ell_L=(\nu_L, e_L)$, the symmetry 
would affect the charged lepton sector, too.
Here, we have required the $2\leftrightarrow 3$ symmetry
only for $\langle \phi_i^u \rangle$, not for 
$\langle \phi_i^d \rangle$, so that the symmetry
does not affect the charged lepton mass matrix.

Of course, the $2\leftrightarrow 3$ symmetry is a phenomenological 
requirement, and the constraint  may be broken.
From the observed constraint \cite{CHOOZ} $(U_\nu)^2 < 0.03$,
we obtain the constraint $|\theta_{\pi\eta}| <12.2^\circ$, i.e.
$$
\left|\frac{v_\pi}{v_\eta}\right| < 0.46 .
\eqno(5.10)
$$

%%%%%%%%%%%%%%%%%%%%%%%%%%%%%%%%%%%%%%%%%%%%%%%%%%%%%%%%%%%%%%%%%%

\vspace{3mm}

{\large\bf 6 \ Concluding remarks}

In conclusion, based on a universal seesaw mass matrix model (1.6) 
with three scalars $\phi_i$, and by assuming an $S_3$ flavor symmetry 
for Yukawa interactions, we have investigated the neutrino masses and mixings. 
For the 
VEV values of $\phi_i^f$ ($f=u,d$), stimulated from a Higgs potential model
\cite{Koide06} for $\phi_i$,  we have assumed the constraint
$$
\langle \phi_\pi^f \rangle^2+\langle \phi_\eta^f \rangle^2=\langle 
\phi_\sigma^f \rangle^2 ,
\eqno(6.1)
$$
where $(\phi_\pi, \phi_\eta, \phi_\sigma)$ are defined by Eq.(1.4). 
However, since we have 4 independent Yukawa coupling constants 
$y_0$, $y_1$, $y_2$ and $y_3$ which are defined by Eq.(2.1),
the model does not have predictability.
Therefore, in the present paper, suggested by the 
observed neutrino mixing (the tribimaximal mixing), 
we have investigated only a simple case with $y_3=0$ where
only the $\nu_\pi$-$\nu_\eta$ mixing is caused.
(Since the observed neutrino mixing is not the exact tribimaximal mixing,
the condition $y_3=0$ is only an approximate requirement for convenience.)
In the case with $y_3=0$ together with the assumption (6.1), 
our conclusion is as follows: the mass eigenvalues depends only 
on the values of the coupling constants $y_0$, $y_1$ 
and $y_2$, while the $\nu_\pi$-$\nu_\eta$ mixing angle $\theta_{\pi\eta}$ 
depends only on the value of $\langle \phi_\pi^u \rangle/\langle \phi_\eta^u 
\rangle$.
Therefore, we can discuss the topic of the neutrino mass spectrum 
independently from that of the deviation from the tribimaximal mixing.

For the neutrino mass spectrum, from the economical point of view of 
the parameter number, we have investigated 
two typical cases with the constraints 
$y_0^2=y_1^2+y_2^2$ and $y_0^2+y_1^2=y_2^2$. 
The former case leads to a case which satisfies Brannen's relation (3.14) for 
the neutrino masses. 
Although the relation (3.14) is very interesting, the theoretical basis of 
the constraint $y_0^2=y_1^2+y_2^2$ is not clear. 
On the other hand, the later case is likely from the view point of the
universality of the coupling constants.
The later case does not satisfy the relation (3.14).
Only for a small value of the parameter $\beta$, the deviation 
from the relation (3.14) can become negligibly small. 
For example, for the solution $\beta=2.94^\circ$ given in Table 1,
the deviation from the relation (3.14) is very small, $\sin^2\beta
=0.003$, as seen in Eq.(4.8), so that the relation (3.14) is approximately 
satisfied.

The neutrino mixing matrix $U_\nu$ can become the tribimaximal mixing (1.1)
in the limit of $v_2^u=v_3^u$ as given in Eq.(5.2).
If we require the $2\leftrightarrow 3$ symmetry for $v_i^u$ (not for $v_i^d$),
we can obtain the tribimaximal mixing (1.1) without affecting the charged
lepton mass spectrum.
Of course, the $2\leftrightarrow 3$ symmetry is a phenomenological 
requirement, and the constraint  may be broken.
From the observed constraint \cite{CHOOZ} $(U_\nu)^2 < 0.03$,
we obtain the constraint $|{v_\pi}/{v_\eta}| < 0.46$.

The numerical predictions of $m_{\nu i}$ shown in Eq.(3.15) and
in Table 1 were obtained by adjusting the parameter $\alpha$
($\beta$) for the observed ratio $\Delta m^2_{solar}/\Delta m^2_{atm}$.
Finally, we would like to give a speculation of neutrino masses
by assuming a simple Yukawa interaction form \cite{Koide0603}
(and without using
the observed value of $\Delta m^2_{solar}/\Delta m^2_{atm}$):
$$
H_\nu=y_\nu \left( \frac{\overline{\ell}_\pi N_\pi
+\overline{\ell}_\eta N_\eta
+\overline{\ell}_\sigma N_\sigma}{\sqrt3}\phi_\sigma^u
+ \frac{\overline{\ell}_\pi N_\eta
+\overline{\ell}_\eta N_\pi}{\sqrt2}\phi_\pi^u
+\frac{\overline{\ell}_\pi N_\pi
-\overline{\ell}_\eta N_\eta}{\sqrt2}\phi_\eta^u \right) .
\eqno(6.2)
$$
Here, in the charged lepton sector, we have assumed 
the universality of the coupling constants on the basis
$( e_1, e_2, e_3)$, while, in the neutrino sector,
we have assumed the universality of those on the S$_3$ 
irrecucible basis
$(\nu_\pi, \nu_\eta, \nu_\sigma)$.
Then, the neutrino masses are predicted as 
$$
\begin{array}{lll}
m_{\nu 1}=\left(\frac{1}{\sqrt6}-\frac{1}{2} \right)^2 m_0^\nu , \\
m_{\nu 2}=\frac{1}{6}m_0^\nu , \\
m_{\nu 3}=\left(\frac{1}{\sqrt6}+\frac{1}{2} \right)^2 m_0^\nu , \\
\end{array}
\eqno(6.3)
$$
without an adjustable parameter. The case predicts
$$
R=\frac{\Delta m_{21}^2}{\Delta m_{32}^2}=\frac{4\sqrt6-9}{4\sqrt6+9}=0.0425 .
\eqno(6.4)
$$
The value (6.4) is somewhat large comparing with the observed value (3.10), 
but, at present, the case is not ruled out within three sigma. Again, 
regarding $m_{\nu 3}$ as $m_{\nu 3}=\sqrt{\Delta m_{atm}^2}$, we predict the 
explicit neutrino mass values as follows:
$$
\begin{array}{lll}
m_{\nu 1}=(5.3_{-0.3}^{+0.4})\times10^{-4}\ {\rm eV}, \\
m_{\nu 2}=(1.05_{-0.05}^{+0.07})\times10^{-2}\ {\rm eV}, \\
m_{\nu 3}=(5.22_{-0.25}^{+0.35})\times10^{-2}\ {\rm eV}.
\end{array}
\eqno(6.5)
$$
The case (6.2) is also interesting because of the simpleness of
its structure.
The predictions (6.4) and (6.5) should be taken as results in
an ideal limit.

In conclusion, the present model (a lepton mass matrix model with
a bilinear form) based on the S$_3$ symmetry has given many interesting
features for the mass spectra and mixings.
However, the model still includes some adjustable parameters.
Further investigation based on another symmetry which gives 
stronger constraints on the parameters than those in the S$_3$
symmetry will be desired.

%%%%%%%%%%%%%%%%%%%%%%%%%%%%%%%%%%%%%%%%%%%%%
\vspace{4mm}
%\newpage

\centerline{\large\bf Acknowledgement} 

This work is supported in part by the Grant-in-Aid for
Scientific Research, Ministry of Education, Science and 
Culture, Japan (No.18540284).

%%%%%%%%%%%%%%%%%%%%%%%%%%%%%%%%%%%%%%%%%%%%
\vspace{6mm}
\centerline{\Large\bf Appendix}

In order to see whether the mass hierarchy is
$m_\eta^2 < m_\sigma^2 < m_\eta^2$ or $m_\pi^2 < m_\eta^2 < m_\sigma^2$,
we estimate the differences among those masses as 
follows:
$$
m_\pi^2-m_\eta^2=\pm\frac{1}{\sqrt3}|y_2|(\sqrt2y_0+y_1)v_u^2 ,
\eqno(A.1)
$$
$$
m_\pi^2-m_\sigma^2=\frac{1}{4\sqrt3}(\sqrt3y_1\pm|y_2|)
(2\sqrt2y_0-y_1\pm\sqrt3|y_2|)v_u^2 ,
\eqno(A.2)
$$
$$
m_\eta^2-m_\sigma^2=\frac{1}{4\sqrt3}(\sqrt3y_1\mp|y_2|)
(2\sqrt2y_0-y_1\mp\sqrt3|y_2|)v_u^2 .
\eqno(A.3)
$$
Since we have defined the factor $(\sqrt2y_0+y_1)$ as positive in Eq.(2.8), 
Eq.(A.1) means that, for the case of the normal hierarchy with 
$m_\pi^2>m_\eta^2$, we must take the upper signs in Eqs.(A.2)-(A.3), i.e.
$$
m_\pi^2-m_\sigma^2=\frac{1}{4\sqrt3}(\sqrt3y_1+|y_2|)
(2\sqrt2y_0-y_1+\sqrt3|y_2|)v_u^2>0 ,
\eqno(A.4)
$$
$$
m_\eta^2-m_\sigma^2=\frac{1}{4\sqrt3}(\sqrt3y_1-|y_2|)
(2\sqrt2y_0-y_1-\sqrt3|y_2|)v_u^2<0 ,
\eqno(A.5)
$$
and, for the case of the inverse hierarchy with $m_\pi^2<m_\eta^2$, 
we must take the lower signs in Eqs.(A.2)-(A.3), i.e.
$$
m_\pi^2-m_\sigma^2=\frac{1}{4\sqrt3}(\sqrt3y_1-|y_2|)
(2\sqrt2y_0-y_1-\sqrt3|y_2|)v_u^2<0 ,
\eqno(A.6)
$$
$$
m_\eta^2-m_\sigma^2=\frac{1}{4\sqrt3}(\sqrt3y_1+|y_2|)
(2\sqrt2y_0-y_1+\sqrt3|y_2|)v_u^2<0 .
\eqno(A.7)
$$

%%%%%%%%%%%%%%%%%%%%%%%%%%%%%%%%%%%%%%%%%%%%

%%%%%%%%%%%%%%%%%%%%%%%%%%%%%%%%%%%%%%%%%%%%%%%%%%%%%%%%%%%%%%%%%%
\newpage

\vspace{3mm}

\centerline{\large\bf Table 1 \ Solutions of $R(\beta)=R_{obs}$}
$$
\begin{array}{|c|ccc|ccc|} \hline
\beta & m_\eta/y_2v_u & m_\sigma/y_2v_u & m_\pi/y_2v_u & 
m_{\nu 1}\ [{\rm eV}] & m_{\nu 2}\ [{\rm eV}] & m_{\nu 3}\ 
[{\rm eV}] \\ \hline
2.94^\circ
&
-0.0775 & 0.3781 & 0.9225
&
0.000368 & 0.00877 & 0.0523 \\[.05in]  
67.59^\circ
&
-0.0775 & -0.3781 & 0.9225
&
0.000368 & 0.00877 & 0.0523 \\[.05in]  
-35.64^\circ
&
0.6636 & 0.6682 & -0.3364
&
0.0515 & 0.0523 & 0.0132 \\[.05in] 
106.17^\circ
&
0.6636 & -0.6682 & -0.3364
&
0.0515 & 0.0523 & 0.0132
\\  \hline
\end{array}
$$

%%%%%%%%%%%%%%%%%%%%%%%%%%%%%%%%%%%%%%%%%%%%%%%%%%%%%%%%%%%%%%%%%%

%%%%%%%%%%%%%%%%%

\end{document}